\documentclass[conference]{IEEEtran}

\IEEEoverridecommandlockouts

\usepackage{amsmath,amsthm,amsfonts}
\usepackage{color}
\usepackage{graphicx}

\usepackage{algorithm}
\usepackage{algpseudocode}

\newcommand{\mbf}[1]{\mathbf{#1}}
\newcommand{\mbb}[1]{\mathbb{#1}}
\newcommand{\mc}[1]{\mathcal{#1}}
\newcommand{\msf}[1]{\mathsf{#1}}

\newcommand{\herm}{\mathsf{H}}
\newcommand{\tran}{\top}
\newcommand{\Uni}{\msf{Uni}}

\newcommand{\ie}{i.e.}

\begin{document}



\title{Intelligent Reflecting Surfaces for Compute-and-Forward} 


\author{\IEEEauthorblockN{Mahdi Jafari Siavoshani$^*$}
\IEEEauthorblockA{\textit{Department of}\\
\textit{Computer Science and Engineering}\\ 
\textit{Sharif University of Technology}\\
Tehran, Iran\\
mjafari@sharif.edu}
\and
\IEEEauthorblockN{Seyed Pooya Shariatpanahi$^*$}
\IEEEauthorblockA{\textit{Electrical and Computer Engineering School}\\
\textit{College of Engineering}\\
\textit{University of Tehran}\\
Tehran, Iran \\
p.shariatpanahi@ut.ac.ir}
\and
\IEEEauthorblockN{Naeimeh Omidvar}
\IEEEauthorblockA{\textit{School of Computer Science} \\
\textit{Institute for}\\
\textit{Research in Fundamental Sciences}\\
Tehran, Iran \\
omidvar@ipm.ir}
}

\maketitle

\bgroup\def\thefootnote{}\footnote{\hspace{-9pt}$^*$These authors have equal contributions to this work.
This work is in part supported by the Iran National Science Foundation (INSF), under the grant number 99022295.}
\egroup

\begin{abstract}
Compute-and-forward is a promising strategy to tackle interference 
and obtain high rates between the transmitting users in a wireless network. However, the quality of the wireless channels between the users substantially limits the achievable computation rate in such systems. 
In this paper, we introduce the idea of  using intelligent reflecting surfaces (IRSs) to enhance the computing capability of the compute-and-forward systems. For this purpose, we consider a multiple access channel (MAC) where a number of users aim to send data to a base station (BS) in a wireless network, where the BS is interested in decoding a linear combination of the data from different users in the corresponding finite field. 
Considering the compute-and-forward framework, we show that through carefully designing the IRS parameters, such a scenario's computation rate can be significantly improved. More specifically, we formulate an optimization problem which aims to maximize the computation rate of the system through optimizing the IRS phase shift parameters. We then propose an alternating optimization (AO) approach to solve the formulated problem with low complexity. Finally, via various numerical results, we  demonstrate the effectiveness of the IRS technology for enhancing the performance of the compute-and-forward systems, which 
indicates 
its great potential for future wireless networks with massive computation requirements, such as 6G.

\end{abstract}


\section{Introduction}
A key technology currently being investigated for 6G is using intelligent reflecting surface (IRS) which can assist communication schemes to arrive at higher performances \cite{Wu2020}. Simply put, by installing large reflecting surfaces -- with adjustable phase shifts on incident waves -- in the wave propagation environment, one can shape the channel's behaviour. Considering the fact that the performance of many communication schemes (especially those related to multiple-input multiple-output MIMO systems) depends heavily on the wireless channel characteristics, this phase shift design flexibility at such surfaces implies great potential performance improvements opportunities \cite{Zhang2020}.    

Accordingly, many wireless communication scenarios have been revisited in terms of such opportunity provided by this new technology, i.e., shaping the channel in favor of the communication scheme. Just to name a few, IRS has been used to assist with improving security at the physical layer such as in \cite{Guan2020}, \cite{Wang2020}, \cite{Chu2020}, and \cite{Lv2020}. Also, the authors in \cite{Ding2020}, \cite{Fang2020}, and \cite{Zeng2020} have considered IRS assisted non-orthogonal-multiple-access (NOMA) scenarios. The papers \cite{Wu2019} and \cite{Zhao2020} investigate the role of IRS in designing well-performing beamformers. Moreover, \cite{Wymeersch2020} and \cite{Liu2020} propose using IRS for wireless localization.

In this paper, we investigate another interesting wireless scenario in which IRS can play a critical role, namely, \emph{physical layer computation}. The main idea in physical layer computation is harnessing interference in the scenarios where only a linear combination, not separate messages, of transmitted data by several users is desired at a single receiver. A well-known approach to implement such an idea is using the compute-and-forward framework \cite{Nazer2011}. Consider a MAC scenario where different users, with different data, wish to communicate their data to a single receiver. However, the receiver is only interested in decoding a linear combination of the messages with given coefficients over a finite field. The proposed scheme in \cite{Nazer2011}, based on nested lattice codes, provides a framework to do this, and furthermore determines the maximum transmission rate of the users so that such computation is feasible, namely, the \emph{computation rate}. 

Following the pioneering work in \cite{Nazer2011}, many research works have benefited from its proposed framework for different communication theory problems (e.g., see \cite{Hong2013}, \cite{Wei2012} and \cite{Lim2020}). 
However, the main limitation of compute-and-forward is the high sensitivity of the computation rate to the wireless channel conditions. More specifically, this framework only allows computing linear combinations with coefficients close to the channel coefficients. Thus, when the receiver is interested in a linear combination which does not match channel coefficients, we face poor performance. This observation is our main motivation for proposing the use of IRS in such a scenario, to alleviate this critical limitation.

Therefore, in this paper we consider the same setup as in \cite{Nazer2011}, equipped with IRS elements installed in the environment (see Fig.~\ref{fig:IRS_CandF_system_model}). Thus, the main problem we address in this paper is how one can tune the phases of IRS elements in order to shape the channel to perform our desired computation. First, this problem is formulated in terms of a non-convex optimization problem, and then by breaking the optimization problem into two simpler sub-problems, we propose an alternating optimization (AO) approach, with good numerical properties, to fine tune the IRS phases. We compare our proposed solution to the random phases adjustments and show the performance improvements achieved.

The most related paper to our work is \cite{Jiang2019}, which proposes IRS-aided over-the-air computation as well. The main difference of our work with \cite{Jiang2019} is that their computation is in the signal domain, while here we perform the computation in the data domain. In other words, the channel domain and the computation domain in \cite{Jiang2019} are both analogue, while here we compute a linear combination of data in a finite field over an underlying analogue channel, which makes our problem more challenging. Thus, their framework and the minimum square error (MSE) approach is not applicable in our setup, and we face a completely different optimization problem. We believe that our framework is more applicable to real-world computation scenarios where almost all calculations should be performed over finite fields in the data domain.

The rest of the paper is organized as follows. In Section \ref{sec:SystemModel}, we explain the system model 
considered in the paper. Section \ref{sec:ProbFormulation} formulates the underlying optimization problem, and Section \ref{sec:Solution} introduces our proposed solution. In Section \ref{sec:Numerical}, we provide numerical results to show the performance improvement of the proposal. Finally, Section \ref{sec:Conclusions} concludes the paper.

\section{System Model}\label{sec:SystemModel}

\begin{figure}
    \centering
    \includegraphics[width=\columnwidth]{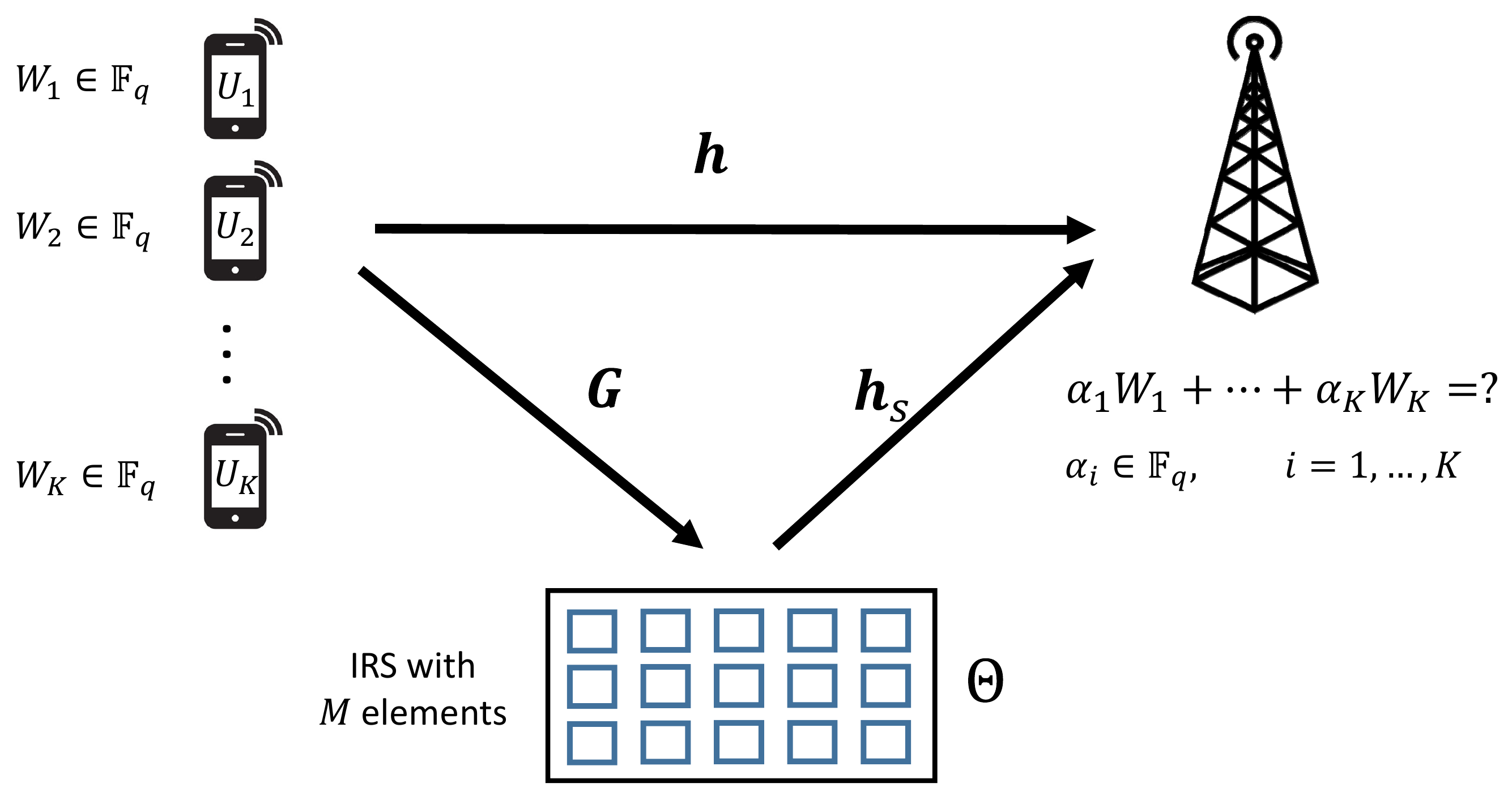}
    \caption{The proposed system model that combines the IRS technique with the compute-and-forward framework.}
    \label{fig:IRS_CandF_system_model}
\end{figure}

As shown in Fig.~\ref{fig:IRS_CandF_system_model}, we consider a system 
consisting of $K$ single-antenna users and one base station (BS), where each user $i$ transmits a file $W_i $ to the BS in the uplink, and the BS is interested in decoding a desired linear combination of users' files, \ie, $\alpha_1W_1+\alpha_2W_2+\cdots+\alpha_KW_K$, where $\alpha_i \in \mathbb{F}_q $ for $i=1,\ldots,K$. We assume that $W_i \in \mathbb{F}_q$, where $\mathbb{F}_q$ is a finite field of size $q$ which is assumed to be a prime number and also all operations are in $\mathbb{F}_q$. 
As introduced in the previous section, to implement such idea, the framework of compute-and-forward comes into the picture, which tries to maximize the transmission rate of the users, known as the computation rate. 

Clearly, the computation rate of the compute-and-forward model highly depends on the channel state information (CSI) between the BS and the users. In this paper and to improve the computation rate of the compute-and-forward approach, we 
introduce an \textit{IRS-assisted compute-and-forward model}, as depicted in Fig.~\ref{fig:IRS_CandF_system_model}, where an IRS consisting of $M$ elements is installed in the environment and helps to enhance the uplink CSI and consequently, the computation rate. 
By assuming a block fading model, and focusing on a single block, the received signal at the BS will be
\begin{equation}
y = \Big( \mbf{h} +  \mbf{G} \mbf{\Theta} \mbf{h}_s \Big)^{\herm} \mbf{x} + z,
\end{equation}
where $\mbf{x}\in\mbb{C}^{K\times 1}$, in which the $i$th element of $\mbf{x}$ denoted by $x_i \in\mbb{C}$ is the signal transmitted by user $i$ with the power constraint $\mathbb{E}(|x_i|^2) \leq \msf{SNR}$, where $\msf{SNR}$ denotes the maximum affordable  average transmit power of each user user, and $(\cdot)^\herm$ denotes the Hermitian operator. Also, $\mathbf{h}=[h_1,\ldots,h_K]^{T}$ represents the direct links of users to the BS collected in the vector $\mathbf{h}$, i.e., $h_i$ is the link from user $U_i$ to BS for $i=1,\ldots,K$, which are assumed to undergo the Rayleigh fading. IRS-related channels include $\mathbf{G} \in\mbb{C}^{K\times M}$, which represents the channel matrix from the users to the IRS with $M$ elements, and $\mathbf{h}_s \in\mbb{C}^{M\times 1}$ which shows the channel vector from the IRS to the BS. All elements of the IRS-related channels are also assumed to undergo Rayleigh fading. Finally, $\mbf{\Theta}=\mathrm{diag}( e^{j\theta_1},\ldots, e^{j\theta_M})$ in which $m$-th IRS element applies a phase shift of $\theta_m \in [0,2\pi]$, and $z\sim \mc{CN}(0,1)$ is the additive Gaussian noise at the BS. The proposed system model is summarized in Fig.~\ref{fig:IRS_CandF_system_model}.

Having the IRS-assisted paths from the users to the BS, we can define the 
effective uplink channel vector as follows
\begin{equation}
    \mbf{h}_{\text{eff}}(\mbf{\Theta}) = \mbf{h} +  \mbf{G} \mbf{\Theta} \mbf{h}_s.
\end{equation}
Therefore, under the compute-and-forward framework, the corresponding computation rate of the considered system can be computed as \cite{Nazer2011}
%
\begin{equation}\label{eq:comp_rate}
    R_{\mathrm{comp}}\big( \mbf{a},\mbf{h}_{\mathrm{eff}}(\mbf{\Theta}) \big) = \log^+ \left( \frac{\msf{SNR}}{\mbf{a}^\herm (\msf{SNR}^{-1} \mbf{I}+ \mbf{h}_{\mathrm{eff}} \mbf{h}_{\mathrm{eff}}^\herm)^{-1} \mbf{a}} \right),
\end{equation}
where the complex integers $\mbf{a}\in \{\mbb{Z} + j\mbb{Z} \}^K$ are some constants determined by the linear combination coefficients $\alpha_1,\ldots,\alpha_K$ in the compute-and-forward framework.


Finally, following many  previous works on IRS systems (e.g., see \cite{Zhang2020}, \cite{Guan2020}, and \cite{Fang2020}), we assume having full CSI of all the channels in the network. We note that although such information may be difficult to attain in practice,  many intelligent solutions have been recently proposed for such challenge (see   \cite{Zhang2020}, \cite{Guan2020}, and \cite{Fang2020} for example), and the research on this issue is still in progress, which is out of the scope of this paper.


\section{Problem Formulation}\label{sec:ProbFormulation}


As aforementioned, in this paper, we aim to maximize the computation rate of the compute-and-forward approach through optimally designing the IRS parameters. 
For this purpose, we try to find the optimal phase shifts at the IRS which maximize the rate given in \eqref{eq:comp_rate}. This problem can be formulated as 
\begin{equation}\label{eq:IRS_Opt_Problem}
\max_{\begin{subarray}{c} \mbf{\Theta}=\mathrm{diag}( e^{j\theta_1},\ldots, e^{j\theta_M}),\\ \forall m\in[1:M]\ : ~\theta_m \in [0,2\pi] \end{subarray}}  R_{\mathrm{comp}}\big(\mbf{a},\mbf{h}_{\mathrm{eff}}(\mbf{\Theta})\big),
\end{equation}
or equivalently, 
\begin{equation}
\max_{\begin{subarray} \mbf{\Theta}=\mathrm{diag}( e^{j\theta_1},\ldots, e^{j\theta_M}),\\ \forall m\in[1:M]\ : ~\theta_m \in [0,2\pi] \end{subarray}}  \log^+ \left( \frac{\msf{SNR}}{\mbf{a}^\herm (\msf{SNR}^{-1} \mbf{I}+ \mbf{h}_{\mathrm{eff}} \mbf{h}_{\mathrm{eff}}^\herm)^{-1} \mbf{a}} \right).
\end{equation}
Alternatively, it can be shown that the above optimization problem can be written as (see \cite{Nazer2011} for the details)
\begin{equation}\label{eq:IRS_Opt_Problem_with x}
\max_{\begin{subarray}{c}\mbf{\Theta}=\mathrm{diag}( e^{j\theta_1},\ldots, e^{j\theta_M}),\\ \forall m\in[1:M]\ : ~\theta_m \in [0,2\pi] \end{subarray}} \max_{\beta \in \mathbb{C} }  \log^+ \left( \frac{\msf{SNR}}{|\beta|^2+\msf{SNR}\|\beta\mbf{h}_\mathrm{eff}(\mbf{\Theta}) - \mbf{a} \|^2} \right)
\end{equation}
where $\beta$ is in fact, the MMSE coefficient at the BS, with which the BS scales its received signal under the compute-and-forward approach \cite{Nazer2011}.





\section{The Proposed Solution}\label{sec:Solution} 

It should be noted that the formulated problem is difficult to solve since it is a non-convex optimization problem. 
In order to efficiently solve this problem, in this section, we propose an alternating optimization (AO) approach that can efficiently address the problem with low complexity. 


The pseudo-code of proposed method is presented in Algorithm \ref{alg:AO_method}. Under the proposed AO method, initially, the IRS is tuned with some initial random phases. Then, in the first step of the algorithm,  treating these phases as fixed and after realizing the CSI of the channels in the network, the effective channel vector of the uplink transmission path is obtained,  
and the objective function in  \eqref{eq:IRS_Opt_Problem_with x} is maximized with respect to the MMSE coefficient, i.e., the control variable $\beta$. In the second step, the updated value of the  MMSE coefficient $\beta$ is considered as fixed and the objective function is maximized with respect to the control variables representing the IRS phase shifts, i.e., $\mbf{\Theta}$. The above steps are done iteratively until convergence. 
Consequently, each iteration of the proposed algorithm contains two optimization sub-problems, which will be elaborated more in the rest of this section.

\begin{algorithm}[b]
\caption{The Proposed Alternating Optimization Algorithm for Optimizing Phase Shifts.} 
\label{alg:AO_method}
\begin{algorithmic}[1]
\Function{CF\_AltOpt}{$\mbf{h}$, $\mbf{G}$, $\mbf{h}_s$, $\mbf{\Theta}_\text{init}$, $\msf{SNR}$, $\mbf{a}$, max\_ao\_itr} \Comment{max\_ao\_itr:~ maximum number of AO iterations}
\State Initialize $\mbf{h}_\text{eff} = \mbf{h} + \mbf{G}\mbf{\Theta}_\text{init}\mbf{h}_s$.
\For{$i\in [1:\text{max\_ao\_itr}]$}
\State Update $\beta \leftarrow \dfrac{\msf{SNR} \mbf{h}_{\text{eff}}^\herm \mbf{a}}{1+\msf{SNR}\|\mbf{h}_{\text{eff}}\|^2}$
\State Find an update of $\mbf{\Theta}$ by performing GD algorithm to minimize $\|\beta \mbf{h}_\mathrm{eff} (\mbf{\Theta}) - \mbf{a}\|^2$.
\State Update $\mbf{h}_\text{eff} = \mbf{h} + \mbf{G}\mbf{\Theta}\mbf{h}_s$. 
\EndFor
\State \Return The achievable computation rate:  

~ $ R_{\text{comp}}=\log^+ \left( \dfrac{\msf{SNR}}{|\beta|^2+\msf{SNR}\|\beta\mbf{h}_\mathrm{eff}(\mbf{\Theta}) - \mbf{a} \|^2} \right)$. 
\EndFunction
\end{algorithmic}
\end{algorithm}

\subsection{Optimizing the MMSE Coefficient $\beta$ Under Fixed IRS Parameters  $\mbf{\Theta}$} 

The first optimization adopts the latest updated IRS phase shift parameters of the IRS  ($\mbf{\Theta}$) 
into the effective channel 
$\mbf{h}_\mathrm{eff}$, and updates the value of the MMSE coefficient $\beta$ 
to maximize the computation rate, i.e., 
\begin{align}\label{eq: P_x}
\mathcal{P}_\beta : \quad     \max_{\beta \in \mathbb{C} }  \log^+ \left( \dfrac{\msf{SNR}}{|\beta|^2+\msf{SNR}\|\beta\mbf{h}_\mathrm{eff}(\mbf{\Theta}) - \mbf{a} \|^2} \right).
\end{align}

Note that since both the logarithmic function and the fractional function $ f(x) = \dfrac{1}{x} $ are monotone, the 
sub-problem $\mathcal{P}_\beta$ can be equivalently written  as the following problem: 
\begin{align}
 \min_{\beta \in \mathbb{C} } \bigg[  |\beta|^2+\msf{SNR}\|\beta\mbf{h}_\mathrm{eff}(\mbf{\Theta}) - \mbf{a} \|^2 \bigg],
\end{align}
which is computationally much easier to solve, since it is just a quadratic optimization. Therefore, to solve this problem and derive the optimal value of $\beta$, it is sufficient to take the derivative of its objective function and put it equal to zero, which results in \cite{Nazer2011} 
\begin{equation}\label{eq:opt_alpha}
    \beta(\mbf{\Theta})=\dfrac{\msf{SNR} (\mbf{h}_\mathrm{eff}^\herm \mbf{a})}{1+\msf{SNR}\|\mbf{h}_\mathrm{eff}\|^2}.
\end{equation}
Consequently, a closed-form expression for the sub-problem $\mathcal{P}_\beta$ is derived, and hence, there is no need to solve an optimization problem in step 1 of each iteration of the proposed algorithm. This significantly saves the computational resources.

\subsection{Optimizing the IRS Parameters $\mbf{\Theta}$ under Fixed MMSE Coefficient $\beta$}  

By fixing  $\beta$ 
to the value derived by \eqref{eq:opt_alpha}, 
the second  optimization tunes the parameters of the IRS to maximize the computation rate $R_{\text{comp}} \big( \mbf{a},\mbf{h}_{\mathrm{eff}}(\mbf{\Theta}) \big)$. 
Therefore, the second sub-problem can be written as
\begin{align}\label{eq: P_theta}
&\mathcal{P}_{\mbf{\Theta}} :\notag\\      
&\max_{ \begin{subarray}{c}\mbf{\Theta}=\mathrm{diag}( e^{j\theta_1},\ldots, e^{j\theta_M}),\\ \forall m\in[1:M]\ : ~\theta_m \in [0,2\pi] \end{subarray} }  \log^+ \left( \frac{\msf{SNR}}{|\beta|^2+\msf{SNR}\|\beta\mbf{h}_\mathrm{eff}(\mbf{\Theta}) - \mbf{a} \|^2} \right).
\end{align}

Similar to the previous sub-problem, it can be verified that this sub-problem is equivalent to the following problem
\begin{align}\label{eq: P_theta equivalent}
 \min_{ \begin{subarray}{c}\mbf{\Theta}=\mathrm{diag}( e^{j\theta_1},\ldots, e^{j\theta_M}),\\ \forall m\in[1:M]\ : ~\theta_m \in [0,2\pi] \end{subarray} }   \|\beta\mbf{h}_\mathrm{eff}(\mbf{\Theta}) - \mbf{a} \|^2,
\end{align}
which is computationally an easier problem to solve compared to the sub-problem $\mathcal{P}_{\mbf{\Theta}}$ itself, since its objective function is a convex quadratic function. 
However, note that this problem is still non-convex, due to the structure of its feasible region over the control variables. To solve this problem, we utilize the gradient descent (GD) approach \cite{boyd2004}, and derive the new values for the IRS parameters $\mbf{\Theta}$. 
Then, the updated IRS parameters are adopted into the effective channel  $\mbf{h}_\mathrm{eff}(\mbf{\Theta})$, which will then be used in the next iteration of the algorithm.


\section{Numerical Analysis}\label{sec:Numerical}

In this section, we numerically evaluate the performance of the proposed IRS-assisted compute-and-forward setup, introduced in Section~\ref{sec:SystemModel}.
Since the optimization problem \eqref{eq:IRS_Opt_Problem} is non-convex, Algorithm~\ref{alg:AO_method} results in a local optimum for each initialization matrix $\mbf{\Theta}_\text{init}$. Hence, in order to grasp a better insight about the performance of the system and the proposed AO-based solution, we run Algorithm~\ref{alg:AO_method} for various channel states (\ie, for different values of $\mbf{h}$, $\mbf{h}_s$, and $\mbf{G}$) and for different initial values of $\mbf{\Theta}_\text{init}$. Then, we find the average of the achievable rate over the initial phases matrix $\mbf{\Theta}_\text{init}$ and also over the channel realizations. 
Algorithm \ref{alg:AO_main_loop} describes the performance evaluation scheme explained above, which finds the average achievable rate of an IRS-assisted compute-and-forward scenario under our proposed method in Algorithm \ref{alg:AO_method}. 



\begin{algorithm}
\caption{The Performance Evaluation Scheme.} 
\label{alg:AO_main_loop}
\begin{algorithmic}[1]
\Require $K$, $M$, $\mbf{a}$, $\msf{SNR}$, num\_chnl\_realz, num\_init\_point 
\State $r$ = array(\text{num\_chnl\_realz}). \Comment{An empty array}
\For{$i\in [1:\text{num\_chnl\_realz}]$}
\State Take random samples of $\mbf{h}$, $\mbf{h}_s$, and $\mbf{G}$ (of proper size, determined by $K$ and $M$) from the Rayleigh distribution.
\State $r_\text{tmp}$ = array(\text{num\_init\_point}). \Comment{An empty array}
\For{$j\in[1:\text{num\_init\_point}]$}
\State Initialize $\mbf{\Theta}_\text{init}$: $\theta_i$'s are i.i.d. and $\theta_i\sim\Uni([0,2\pi])$.
\State $r_\text{tmp}[j] = \text{CF\_AltOpt}(\mbf{h}, \mbf{G}, \mbf{h}_s, \mbf{\Theta}_\text{init}, \msf{SNR}, \mbf{a},$\newline 
$\text{max\_ao\_itr})$.
\EndFor
\State $r[i] = \text{average}_\ell \left( r_\text{tmp}[\ell] \right)$.  \Comment{Finds the average rate over the different initializations of $\mbf{\Theta}_{\text{init}}$} \label{alg:line:avg_over_init_phases}
\EndFor
\State \Return $\text{average}_\ell \left( r[\ell] \right)$. \Comment{Finds the average rate over the different channel realizations}
\end{algorithmic}
\end{algorithm}

\subsection{Baselines}

To better investigate the proposed IRS-assisted compute-and-forward method, we compare the average achievable rate of the AO method, obtained by Algorithm~\ref{alg:AO_main_loop}, with some other baseline algorithms, described in the following. 

\textbf{No-IRS Baseline:} This baseline assumes there are no IRS elements installed in the environment (\ie, $M=0$), so we have a plain compute-and-forward scenario. In this case the achievable compute-and-forward rate can be simply derived by \eqref{eq:comp_rate}, where $\mbf{h}_{\text{eff}} = \mbf{h}$. 

\textbf{RndPhz-avg Baseline (random phase shifts with averaging):} As the simplest baseline for the IRS-assisted compute-and-froward scenario, we use the rate achieved by choosing the phase shift  matrix $\mbf{\Theta}$ randomly such that $\theta_i\sim\Uni([0,2\pi])$, and then average the results over both the channel realizations and the random phase shift matrix $\mbf{\Theta}$. 
 
 \textbf{RndPhz-max Baseline (random phase shifts with maximizing):} This baseline is the same as RndPhz-avg, but the results is obtained by taking the maximum over the rates achieved for each phase shift matrix $\mbf{\Theta}$, and then taking the average over the channel realizations. 

\textbf{AO-max Baseline (AO with maximizing  over the initial phase shifts):} This baseline is similar to Algorithm~\ref{alg:AO_main_loop}, but the Line~\ref{alg:line:avg_over_init_phases} is replaced with finding the maximum over the initial phase shift matrix $\mbf{\Theta}_{\text{init}}$.


\subsection{Comparison with the Plain Compute-and-Forward Scheme}
In order to show the benefit of utilizing IRS in a compute-and-forward system, here we compare the performance of the proposed IRS-assisted approach with the performance of the plain compute-and-forward method (i.e., without IRS), where $\mbf{h}_{\text{eff}} = \mbf{h}$. 

\begin{figure}
    \centering
    \includegraphics[width=\columnwidth]{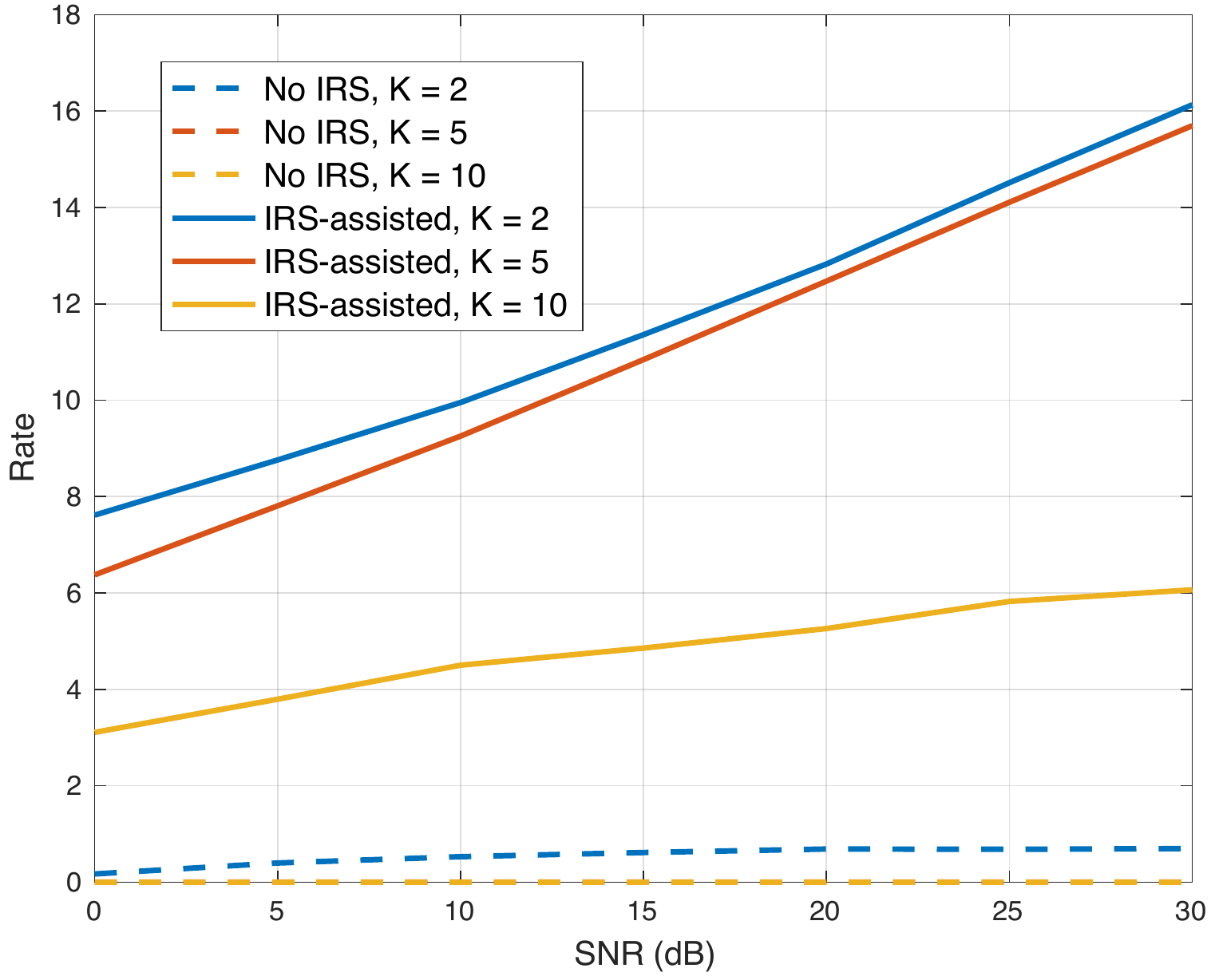}
    \caption{The achievable computation rate versus SNR for No-IRS and the proposed IRS-assisted scenarios. The results are averaged over the channel realizations and the initial phase shifts. For the number of IRS elements in the IRS-assisted case, we have $M=20$. In each case, the receiver is aimed to decode the sum of transmitted signals, namely, $\mathbf{a}=[1 \cdots 1]^\tran$.}
    \label{fig:rate_vs_snr}
\end{figure}

In Fig.~\ref{fig:rate_vs_snr}, the achievable rate of the proposed Algorithm~2 is presented versus $\msf{SNR}$, for different number of users $K$ (here, we fix the number of IRS elements to $M=0$, \ie, No-IRS baseline and $M=20$, \ie, the proposed IRS-assisted scenario). As it is observed in Fig.~\ref{fig:rate_vs_snr}, the proposed IRS-assisted setup significantly outperforms the No-IRS scenario. In fact for the cases of $K=2$ and $K=5$, the No-IRS scenario, on average, achieves almost zero rate.
Moreover, we can observe from Fig.~\ref{fig:rate_vs_snr} that computing a linear combination of data from users becomes a more challenging task as the number of users grows, which is reflected in the lower computation rate.

\subsection{Comparison with the Other IRS-Assisted Baselines}
\begin{figure}
    \centering
    \includegraphics[width=\columnwidth]{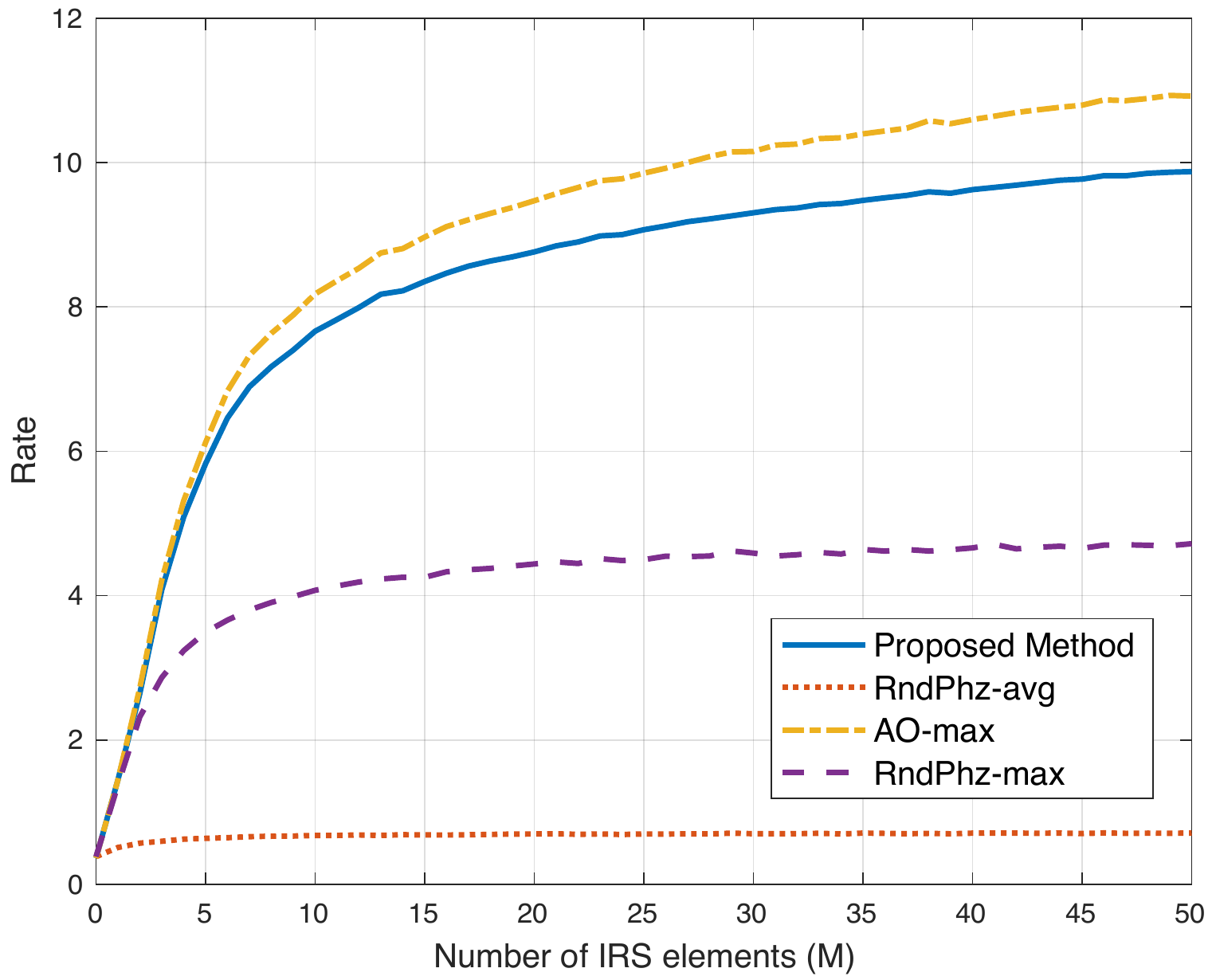}
    \caption{The achievable computation rate versus the number of IRS elements, \ie, $M$, is presented for the proposed method, derived using Algorithm~\ref{alg:AO_main_loop}, and compared with the other baselines. Here, we assume $K=2$, $\msf{SNR}=5\mathrm{dB}$, and $\mathbf{a}=[1\ \ 1]^\tran$.}
    \label{fig:rate_vs_M}
\end{figure}

The results of comparing Algorithm~\ref{alg:AO_main_loop} with RndPhz-avg, RndPhz-max, and AO-max baselines can be found in Fig.~\ref{fig:rate_vs_M},  where the achievable computation rates are depicted versus the number of IRS elements $M$. For each method and for each set of problem parameters (\ie, each point of Fig.~\ref{fig:rate_vs_M}), we take a number of independent channel realization $\mathrm{num\_chnl\_realz}$ (which is $350$ for the proposed approach and AO-max, and $5350$ for the RndPhz-avg and RndPhz-max algorithms), and for each realization, we choose $\mathrm{num\_init\_point}=35$ random initial phase shifts. 
Moreover, in Fig~\ref{fig:rate_vs_M}, we assume the number of transmitters is $K=2$, $\msf{SNR}=5\mathrm{dB}$, and the base station is interested in decoding the sum of transmitted symbols (over the finite field $\mathbb{F}_q$), namely, we choose $\mathbf{a}=[1\ \ 1]^\tran$. As Fig.~\ref{fig:rate_vs_M} shows, the proposed AO approach substantially improves the computation rate compared to the random phase shift selection scheme (\ie, the RndPhz-avg and RndPhz-max baselines). At the same time, the performance of the proposed approach is not far from the AO-max baseline, which has a much higher computation complexity and is not practical in real-world scenarios.

\subsection{No Direct Link Scenario}
\begin{figure}
    \centering
    \includegraphics[width=\columnwidth]{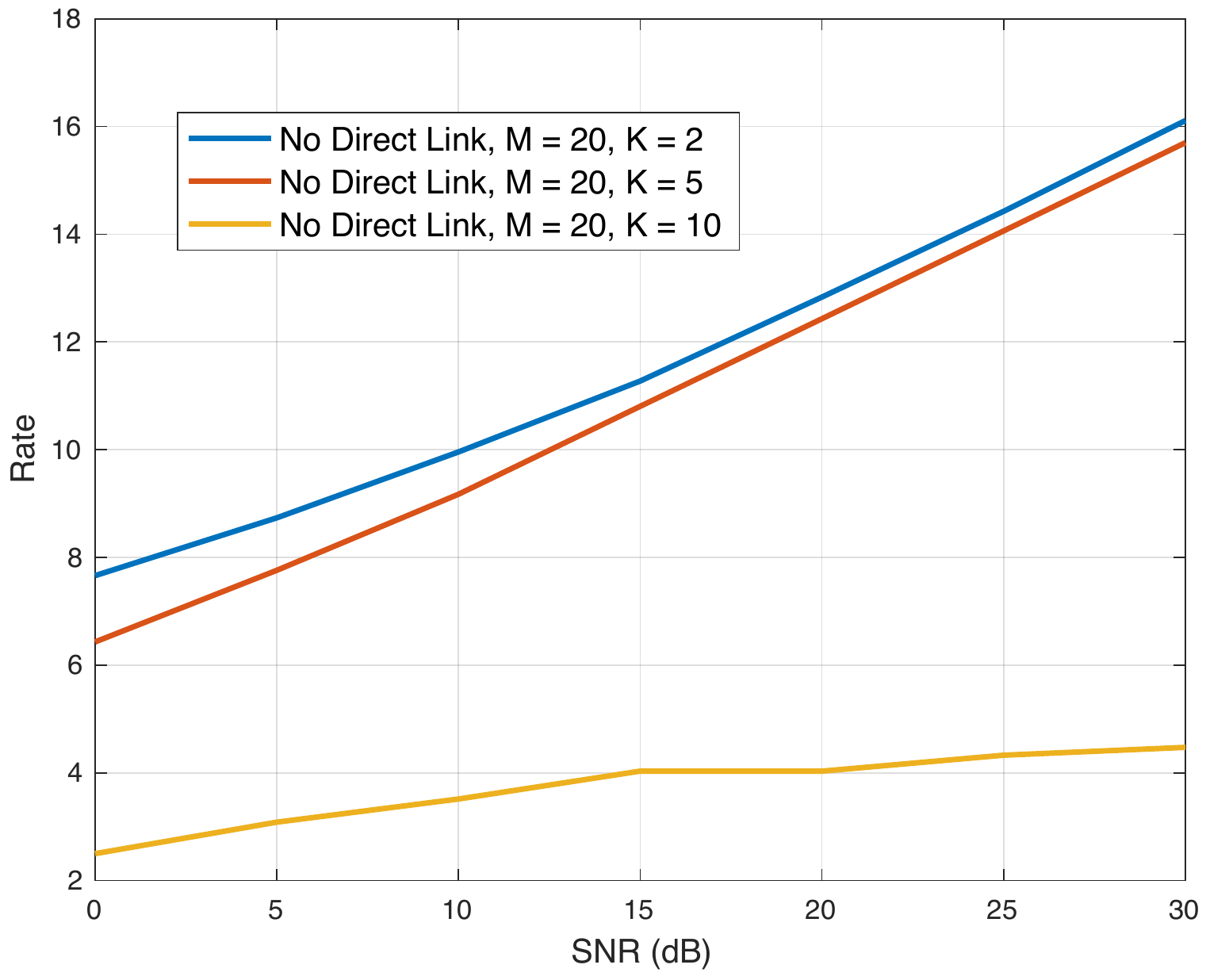}
    \caption{The achievable computation rate versus SNR is presented for the proposed approach, derived using Algorithm~\ref{alg:AO_main_loop}, where no direct links are present between the users and the base station (\ie, $\mbf{h}=0$). Here, for the number of IRS elements, we have $M=20$. In each case, the receiver is interested in the sum of transmitted signals, namely, $\mathbf{a}=[1 \cdots 1]^\tran$.}
    \label{fig:rate_vs_snr-NDL}
\end{figure}

\begin{figure}
    \centering
    \includegraphics[width=\columnwidth]{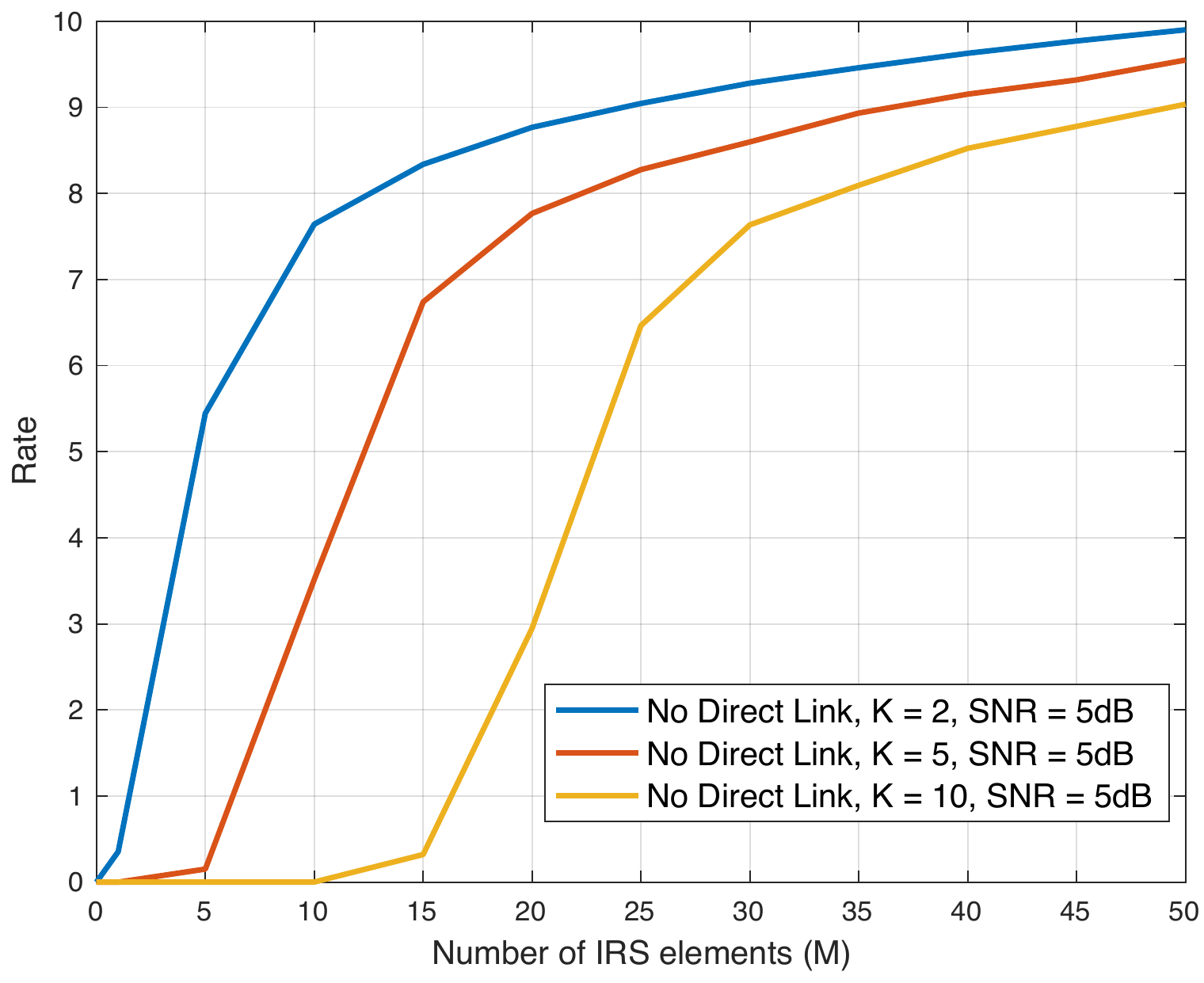}
    \caption{The achievable computation rate versus the number of IRS elements $M$ is presented for the proposed approach, derived using Algorithm~\ref{alg:AO_main_loop}. Here, we assume no direct link between the users and the base station (\ie, $\mbf{h}=0$), $\msf{SNR}=5\mathrm{dB}$, and $\mathbf{a}=[1 \cdots 1]^\tran$.}
    \label{fig:rate_vs_M-NDL}
\end{figure}

In many applications, there may be cases where no direct uplinks are available to the users. In such cases, the IRS deployment can be significantly beneficial as demonstrated in Figs.~\ref{fig:rate_vs_M-NDL}~and~\ref{fig:rate_vs_snr-NDL}.

Figs.~\ref{fig:rate_vs_M-NDL}~and~\ref{fig:rate_vs_snr-NDL} depict similar scenarios as in Figs.~\ref{fig:rate_vs_snr}~and~\ref{fig:rate_vs_M}, but for the case where there are no direct links between the users and the base station, \ie, $\mathbf{h}=0$. The general behavior of the performance versus the network parameters is the same, which suggests that the link involving IRS is the main player determining the performance.

\section{Concluding Remarks}\label{sec:Conclusions}
In this paper, we have proposed employing IRS technology to enhance the rate of computing a linear combination of distributed data among different mobile devices, at a base station. In order to do this, we have used the well-known framework of compute-and-forward, which has enabled us to perform the computation in the data domain, \ie, the corresponding finite field.
Our proposal includes an alternating optimization approach to tune the IRS elements' phase shifts in order to maximize the computation rate, which can alternatively be interpreted as the \emph{IRS's computation power}. Our numerical results demonstrate the great potential of using intelligent reflecting surfaces for various application in 
edge computing scenarios (e.g., the computation tasks in federated learning applications for next generation communication networks).

\ifCLASSOPTIONcaptionsoff
  \newpage
\fi

\end{document}